\documentclass[superscriptaddress, amsmath, amssymb, aps,pra,nofootinbib,twocolumn
]{revtex4-2}
\usepackage{graphicx} 
\usepackage{comment}
\usepackage{dsfont}
\usepackage{color}
\usepackage{bm} 
\usepackage{hyperref}

\usepackage[utf8]{inputenc}
\usepackage{amsmath}
\usepackage{amssymb}
\usepackage[a4paper, total={6.6in, 8.8in}]{geometry}
\usepackage{float}
\usepackage{braket}
\usepackage{adjustbox}
\usepackage{color}
\usepackage{framed} 
\usepackage{url}
\usepackage[normalem]{ulem}
\usepackage{subfigure}

\begin{document}

\title{A Toffoli Gadget for Magnetic Tunnel Junctions Boltzmann Machines}

\author{Dairong Chen}
\affiliation{Center for Quantum Phenomena, Department of Physics, New York University, New York, NY 10003 USA}

\author{Augustin Couton Wyporek}
\affiliation{Center for Quantum Phenomena, Department of Physics, New York University, New York, NY 10003 USA}
\affiliation{Université de Lorraine, CNRS, Institut Jean Lamour, F-54000 Nancy, France}

\author{Pierre Chailloleau}
\affiliation{Université de Lorraine, CNRS, Institut Jean Lamour, F-54000 Nancy, France}

\author{Ahmed Sidi El Valli}
\affiliation{Center for Quantum Phenomena, Department of Physics, New York University, New York, NY 10003 USA}
\author{Flaviano Morone}
\affiliation{Center for Quantum Phenomena, Department of Physics, New York University, New York, NY 10003 USA}

\author{Stephane Mangin}
\affiliation{Université de Lorraine, CNRS, Institut Jean Lamour, F-54000 Nancy, France}

\author{Jonathan Z. Sun}
\affiliation{IBM T. J. Watson Research Center, Yorktown Heights, NY 10598, USA}

\author{Dries Sels}
\affiliation{Center for Quantum Phenomena, Department of Physics, New York University, New York, NY 10003 USA}
\affiliation{Center for Computational Quantum Physics, Flatiron Institute, New York, NY, USA}

\author{Andrew D. Kent}
\affiliation{Center for Quantum Phenomena, Department of Physics, New York University, New York, NY 10003 USA}

\begin{abstract}
Magnetic Tunnel Junctions (MTJs) are of great interest for non-conventional computing applications. The Toffoli gate is a universal reversible logic gate, enabling the construction of arbitrary boolean circuits. Here, we present a proof-of-concept construction of a gadget which encodes the Toffoli gate's truth table into the ground state of coupled uniaxial nanomagnets that could form the free layers of perpendicularly magnetized MTJs. This construction has three input bits, three output bits, and one ancilla bit. We numerically simulate the seven macrospins evolving under the stochastic Landau–Lifshitz–Gilbert (s-LLG) equation. We investigate the effect of the anisotropy-to-exchange-coupling strength ratio $H_A/H_\text{ex}$ on the working of the gadget. We find that for $H_A/H_\text{ex} \lesssim 0.93$, the spins evolve to the Toffoli gate truth table configurations under LLG dynamics alone, while higher $H_A/H_\text{ex}$ ratios require thermal annealing due to suboptimal metastable states. Under our chosen annealing procedure, the s-LLG simulation with thermal annealing achieves a 100\% success rate up to $H_A/H_\text{ex}\simeq3.0$. The feasibility of constructing MTJ-free-layer-based Toffoli gates highlights their potential in designing new types of MTJ-based circuits.

\end{abstract}

\maketitle

\section{\label{sec:intro}Introduction}

Magnetic tunnel junctions (MTJs) hold significant potential in advancing future computing technologies. Traditionally, MTJs have been employed in the development of long-term storage elements and magnetic random access memory (MRAM). These applications utilize the stable and non-volatile characteristics of MTJs to store data efficiently and reliably~\cite{Kent2015,ender2021emerging}. On the other hand, the stochastic behavior and dynamic nature of MTJs have led to novel computing applications beyond data storage~\cite{finocchio_roadmap_2024}. With the rapid advances in machine learning and artificial intelligence, neuromorphic computing has emerged as a promising approach to meet the increasing demand for hardware capable of handling large-scale data processing, resolving the memory bottleneck issue, and having less energy consumption \cite{schuman_opportunities_2022,zhang_neuro-inspired_2020}. MTJs have been shown to be a prominent candidate in the development of neuromorphic computing hardware~\cite{torrejon_neuromorphic_2017,romera_vowel_2018,grollier_neuromorphic_2020,markovic_physics_2020}. 

MTJs also provide novel avenues for solving optimization problems. Probabilistic Ising machines built using MTJs have been used to find approximate solutions to NP-hard problems by encoding the true solution in the system's ground state~\cite{camsari_stochastic_2017,borders_integer_2019,si_energy-efficient_2024}. Recently, several of the authors of this article conducted numerical studies to benchmark the performance of Landau-Lifshitz-Gilbert (LLG) dynamical systems in solving combinatorial optimization problems~\cite{chen2024solvingcombinatorialoptimizationproblems}. This work demonstrated that the stochastic and nonlinear dynamics governed by the stochastic LLG equation has the potential to find better solutions to NP problems as system size scales up~\cite{chen2024solvingcombinatorialoptimizationproblems}.

\begin{figure}
 \centering
 \includegraphics[width=0.5\textwidth]{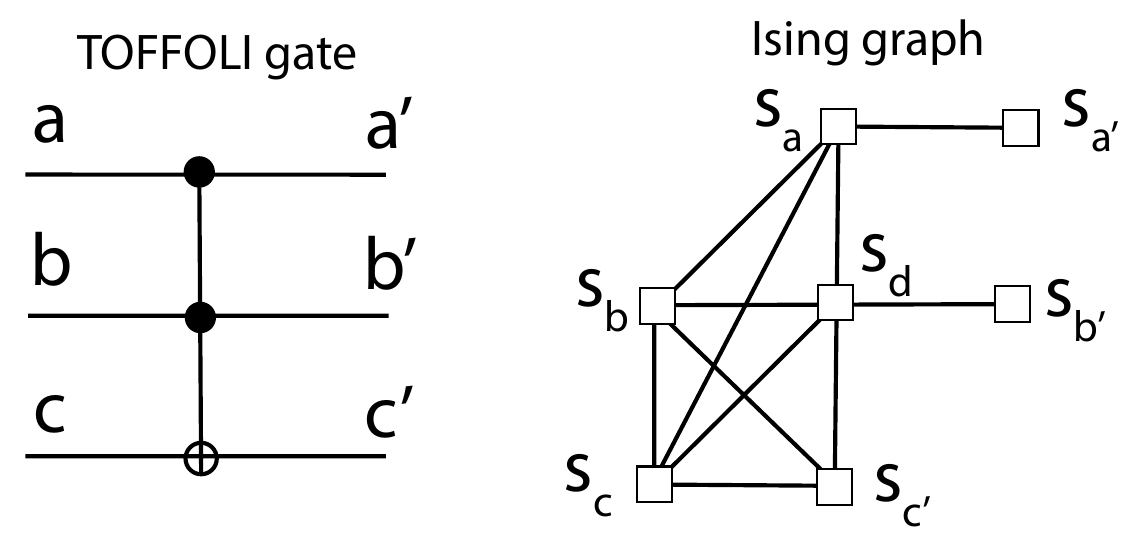}
 \caption{The circuit representation of Toffoli gate (left) and an Ising spins network for the construction of a Toffoli gate (right)~\cite{chamon_quantum_2017}.} 
 \label{fig:ising graph}
\end{figure}

Building on these insights, we aim to broaden the scope of these unconventional computing application of MTJs. Although some combinatorial optimization problems, such as 3-SAT, are NP-complete and thus universal for computation, it can be quite cumbersome to encode the problem of interest into one of these Boolean satisfiability problems. It would be useful to be able to efficiently encode simple tasks, such as addition, multiplication or other basic operations into these unconventional dynamical systems. On the most general level, it suffices to encode a universal (reversible) logical gate into the ground state of some coupled MTJs. Here we focus on the Toffoli gate, which is a universal reversible gate~\cite{toffoli_reversible_1980} that can serve as the building block for arithmetic operations. In the field of magnetism, efforts have been made to develop the Toffoli gate using three interacting classical spins~\cite{Nuzzi2024}, skyrmions~\cite{costilla_implementation_2023,chauwin_skyrmion_2019}, and spin waves~\cite{balynskiy_reversible_2018}. In this paper, we model the construction of a classical Toffoli gate using seven coupled uniaxial nanomagnets, which could represent the free layers of seven perpendicularly magnetized MTJs. We use the macrospin approximation for each free layer and apply LLG dynamics to simulate their evolution.

In the following sections, we will discuss our methods and results in detail. We begin by describing the mapping of the Toffoli gate's truth table to the ground state of an Ising Hamiltonian. Next, we outline our approach to simulating the macrospin model using the stochastic Landau-Lifshitz-Gilbert (sLLG) equation and discuss how macrospins can emulate Ising spins. We then detail our method for using coupled macrospins to construct a Toffoli gate. Finally, we present the results of our macrospin Toffoli gate simulations, both at zero temperature and using simulated annealing.

\section{\label{sec:methods}Methods}

\subsection{\label{sec:toffoli_construct}Toffoli Gate Construction}
The truth table of the Toffoli gate is described by the following Boolean function:
\begin{equation}
\{a,b,c\} \rightarrow \{a,b, c \,\oplus (a \,\wedge \, b)\}.
\end{equation}
That is, the last bit flips conditional on the first two bits being 1. (It can be considered a `2'-controlled-NOT gate.) This gate is universal, as $\{a,b,1\}$ will result in NAND of the inputs $a$ and $b$, \emph{i.e.}, $\{a,b, \neg (a \, \wedge \, b) \}$ and the NAND gate is a universal digital logic gate. Since the truth table can be represented as a permutation matrix, the Toffoli gate is also reversible.

This truth table of the Toffoli gate can be mapped to the ground states of a seven-coupled Ising spin system as shown in Chamon {\em et al.}~\cite{chamon_quantum_2017}. We denote the Ising spins as $s$, which can only take two values $s=\pm1$. The connection network between Ising spins is depicted in Fig.~\ref{fig:ising graph}. In this system, we denote the spins that act as inputs $s_a$, $s_b$, $s_c$, the spin that acts as an ancilla $s_d$, and the spins that act as the outputs as $s_{a'}$, $s_{b'}$, $s_{c'}$. The Hamiltonian for this Toffoli gate construction is given by~\cite{chamon_quantum_2017} 
\begin{equation}
    H_T = -\sum_{\langle i,j \rangle} J_{ij} s_i s_j - \sum_i h_i s_i,
    \label{eq:H_Toffoli}
\end{equation}
where the $J_{ij}$ is the coupling term between spins $i$ and $j$, and $h_i$ is the local field term applied to each individual spin. The values for $J_{ij}$ and $h_i$ are shown in Table~\ref{table:Jcoupling}.

\begin{table}[htbp]
\centering
\begin{tabular}{||c c c | c | c c c|| c ||} 
 \hline
 $s_a$ & $s_b$ & $s_c$ & $s_d$ & $s_{a'}$ & $s_{b'}$ & $s_{c'}$ & \\ 
 \hline\hline
 -1 & 3 & 2 & -4 & 1 & 0 & -2 & $s_a$   \\ 
 3 & 3 & -4 & 8 & 0 & 1 & 4 & $s_b$    \\ 
  2 & -4 & 2 & 6 & 0 & 0 & 4 & $s_c$    \\ 
  \hline
   -4 & 8 & 6 & -4 & 0 & 0 & -6 & $s_d$    \\ 
   \hline
   1 & 0 & 0 & 0 & 0 & 0 & 0 & $s_{a'}$   \\ 
0 & 1 & 0 & 0 & 0 & 0 & 0 & $s_{b'}$  \\ 
-2 & 4 & 4 & -6 & 0 & 0 & -2 & $s_{c'}$  \\  [1ex]
 \hline
\end{tabular}
\caption{$J_{ij}$-coupling between the input spins $\{s_a,s_b,s_c\}$, output $\{s_{a'},s_{b'},s_{c'}\}$ and the ancilla $s_d$. The diagonal represents the external field $h_i$ applied on the $i^\mathrm{th}$ spin.}
\label{table:Jcoupling}
\end{table}

\begin{figure}
 \centering
 \includegraphics[width=0.48\textwidth]{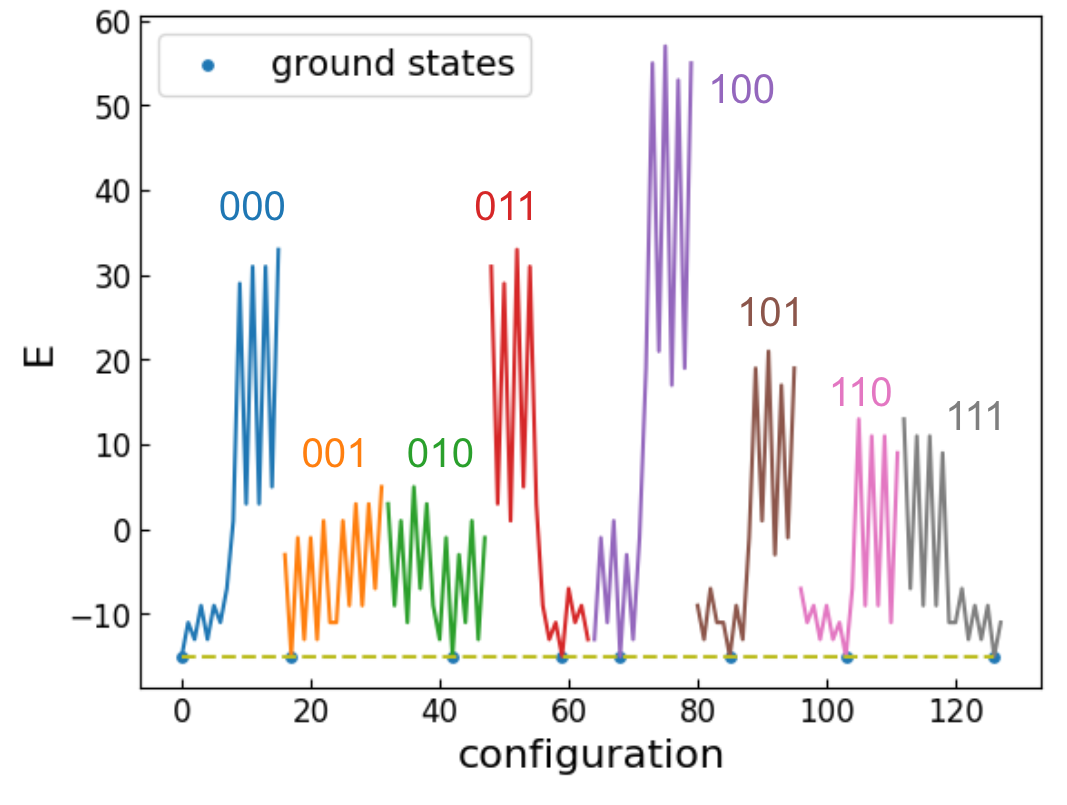}
  \caption{The energy landscape calculated from the Hamiltonian $H_\text{T}$. The x axis represents the decimal representation of the  binary string $(a,b,c,d,a',b',c')$ and the y axis is the energy calculated from $H_T$ by converting each logical bits to its corresponding Ising spins. Each color represents a set of configurations with the same fixed input values $(a,b,c)$. The blue dots represent the eight configurations with the lowest energies, which have the correct input-output relation according to the truth table for the Toffoli gate. The unit of energy will be taken to be $k_BT_\text{amb}$, where $k_B$ is the Boltzmann constant and $T_\text{amb}=300$ K, in the stochastic macrospin model discussed in Secs.~\ref{sec:toffoli_implement} and~\ref{sec:simulated_anneal}.}
   \label{fig:energy landscape}
\end{figure}

The Ising spins can only take two values: $s=\pm 1$, where $s=+1$ corresponds to logical 1, and $s=-1$ corresponds to logical 0. We denote logical input as $a$, $b$, $c$, logical ancilla as $d$, and logical output as $a'$, $b'$, $c'$. For simplicity, the configuration of the system is denoted as a binary string $(a,b,c,d,a',b',c')$, and we label each configuration by the decimal value of this binary string. For example, an input $001$, an ancilla $0$, and an output $001$ form the binary string $(0,0,1,0,0,0,1)$, which corresponds to $17$ in decimal representation. With this convention, we plot the energy landscape of the Hamiltonian $H_T$ for different configurations in Fig.~\ref{fig:energy landscape}. The configurations with the lowest energies correspond to the correct input-output relations for the Toffoli gate's truth table.

Given the Toffoli gate Hamiltonian $H_T$, we model a system of 7 coupled single-domain ferromagnets that represent the 7 Ising spins. The macrospin approximation is applied to each domain such that the magnetization of the spins in the $i^\mathrm{th}$ domain are represented by one giant magnetic moment $\Vec{m}_i = m_i\cdot\hat{m}_i$, with magnitude $m_i$ and unit direction vector $\hat{m}_i$. This giant magnetic moment $\Vec{m}_i$ is referred to as a macrospin. We denote the macrospin system as $(\Vec{m}_a,\Vec{m}_b,\Vec{m}_c,\Vec{m}_d,\Vec{m}_{a'},\Vec{m}_{b'},\Vec{m}_{c'})$. In contrast to Ising spins that are either up or down, a macrospin $\Vec{m}$ is defined on the surface of a 3D sphere with two degrees of freedom (in spherical coordinates, $\theta$ and $\phi$), and evolves according to the LLG equation.

\begin{figure}
 \centering
 \includegraphics[width=0.45\textwidth]{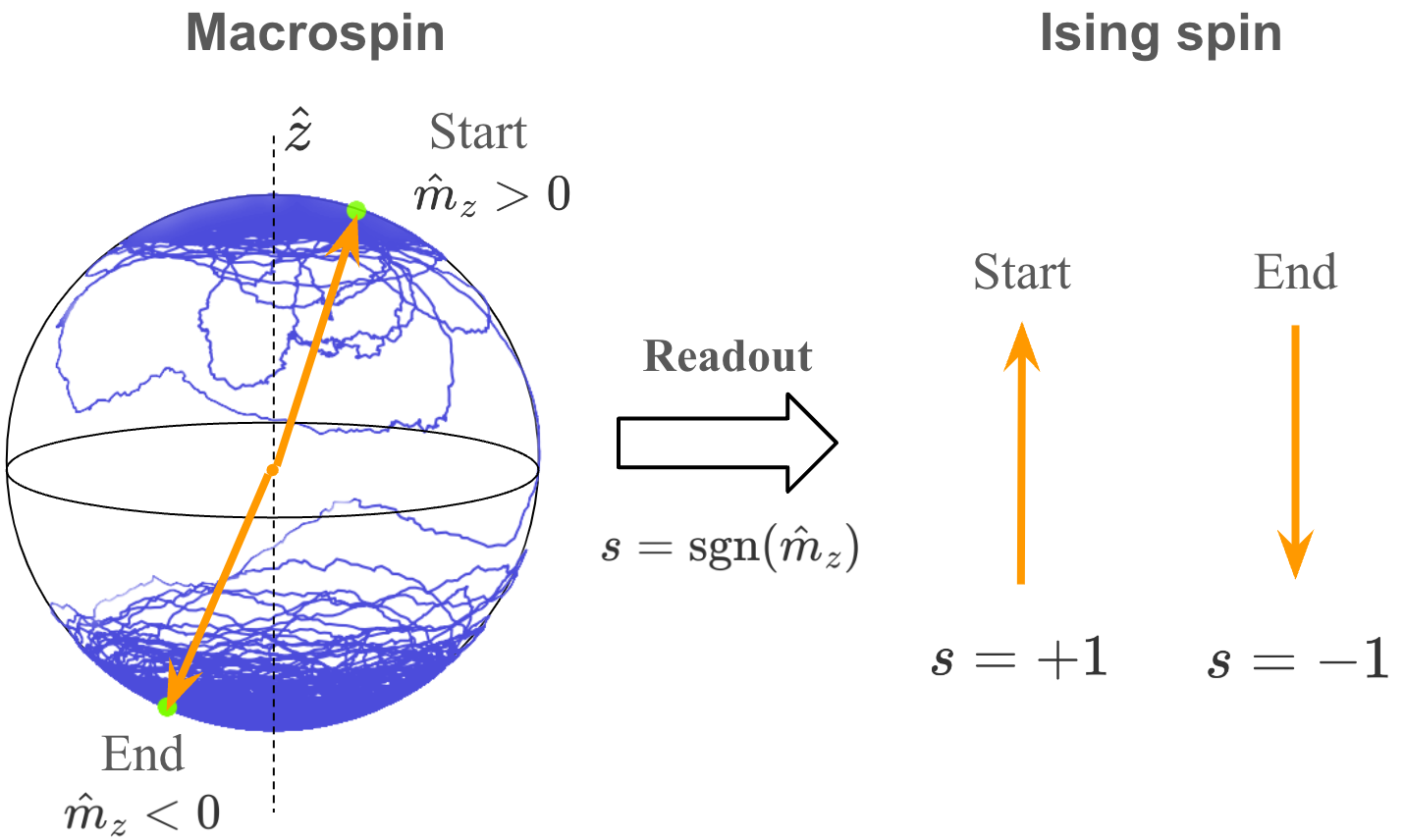}
 \caption{Illustration of macrospin dynamics (left) and its maping to an Ising spin (right).} 
 \label{fig:ising and macro}
\end{figure}

To emulate Ising spins using macrospins that follow full classical LLG dynamics, we include a uniaxial magnetic anisotropy along the $\hat{z}$-axis, such that the macrospins favor aligning up or down along the $\hat{z}$-axis. We initialize the system by converting $s=\pm1$ to $\Vec{m} = \pm m \cdot \hat{z}$, aligning the macrospin with the north or south pole based on the sign of the Ising spin. The three macrospins that act as the input of the Toffoli gate, $(\Vec{m}_a,\Vec{m}_b,\Vec{m}_c)$, are fixed numerically to their initial states throughout the simulation, while the other four macrospins $(\Vec{m}_d,\Vec{m}_{a'},\Vec{m}_{b'},\Vec{m}_{c'})$ undergo LLG evolution. For readout at the end of the simulation, the sign of the projection of the macrospin's moment on the z-axis ($\mathrm{sgn}(\hat{m_z})$) is the binary value of Ising spin, as illustrated in Fig.~\ref{fig:ising and macro}.

\subsection{\label{sec:LLG}LLG and s-LLG Equations}
To describe the LLG dynamics that each macrospin follws, we begin by considering the micromagnetic energy of the system, $E$, which consists of three contributions:
\begin{align}
    E = E_\text{ex} + E_\text{anis} + E_\text{Z},
    \label{eq:micromag energy}
\end{align}
where $E_\text{ex}$ is the exchange energy between each pair of moments, $E_\text{anis}$ is the anisotropic energy arising from the 
crystallographic structure of the domain, and $E_\text{Z}$ is the Zeeman energy from an external field applied to the moment. The expression for each term is given by:
\begin{align}
    E_\text{ex} &= -\mu_0 \sum_{\langle i,j \rangle}^{n} 2 J_{ij} H_\text{ex}\frac{\Vec{m}_i \cdot \Vec{m}_j}{m_i+m_j},\label{eq:E_ex} \\
    E_\text{anis} &= -\frac{1}{2}\mu_0 \sum_i^n H_{Ai} m_i (\hat{e}_{hi} \cdot \hat{m}_i)^2, \label{eq:E_anis} \\
    E_\text{Z} &=-\mu_0 \sum_i^n \Vec{m}_i\cdot h_i H_\text{ext}\hat{e}_{\text{ext}i}, 
    \label{eq:E_Z}
\end{align}
where $\mu_0$ is the vacuum permeability, $J_{ij}$ are dimensionless coupling terms characterizing the exchange interaction between magnetic moments, and $H_\text{ex}$ (with units of field) characterizes the general scale of the exchange coupling strength. The summation over $\langle i,j \rangle$ is over pairs of connected spins. $H_{Ai}$ characterizes the magnetic anisotropy of moment $i$, and $\hat{e}_{hi}$ is the unit vector representing the preferred crystallographic direction of moment $i$, which we set $\hat{e}_{hi} = \hat{z}$ to facilitate readout along this axis. $h_i$ are dimensionless terms that characterize the external magnetic field applied to each magnetic moment, and $H_\text{ext}$ (with units of field) characterize the general scale of the external field strength. $\hat{e}_{\text{ext}i}$ is the unit vector in the direction of the applied external field. We set $\hat{e}_{\text{ext}i}=\hat{z}$, so that this field is along the same axis as the magnetic anisotropy.

The $i^\mathrm{th}$ magnetic moment experiences an effective field $\Vec{H}_{\text{eff},i}$, which can be obtained by computing the 
derivative of the micromagnetic energy $E$ with respect to the magnetic moment $\Vec{m}_i$: 
\begin{align}
    \Vec{H}_{\text{eff},i} = -\frac{1}{\mu_0}\frac{\partial E}{\partial \Vec{m}_i}.
    \label{eq:Heff}
\end{align}
The expression for $\Vec{H}_{\text{eff},i}$ is substituted into LLG equation, which determines the dynamical evolution of each magnetic moment at each time $t$:
\begin{align}
    \frac{d\Vec{m}_i}{dt} = \gamma' \Vec{m}_i \times \Vec{H}_{\text{eff},i} - \frac{\alpha_i}{m_i} \Vec{m}_i \times  \frac{d\Vec{m}_i}{dt},
    \label{eq:LLG}
\end{align}
where $\gamma'= \mu_0 \gamma$, $\gamma$ is the gyromagnetic ratio. $\alpha_i$ is the damping constant of the $i^\mathrm{th}$ moment. 

To account for the stochastic nature of the system under thermal fluctuations, a Langevin field term $\Vec{H}_{\text{th},i}$ is added to the effective field $\Vec{H}_{\text{eff},i}$:
\begin{align}
    \Vec{H}_{\text{eff},i} \rightarrow \Vec{H}_{\text{eff},i} + \Vec{H}_{\text{th},i}.
\end{align}
The Langevin fields at times $t$ and $t'$ follow the statistical properties described in Refs.~\cite{brown_thermal_1963,berkov_magnetization_2007,GarciaLazaro98}:
\begin{gather}
    \langle \Vec{H}_{\text{th},i} \rangle = 0, \\
    \label{eq:Brown_space_time} 
    \langle \Vec{H}_{\text{th},i,m}(t)\Vec{H}_{\text{th},i,n}(t')\rangle = C_i \delta_{mn} \delta (t-t'), \\
    C_i = \frac{2\alpha_i k_B T}{\mu_0^2 m_i \gamma(1+\alpha_i^2)},
\end{gather}
where $m$ and $n$ represent the $x, y, z$ components of the random field, and $\delta_{mn}$ and $\delta(t-t')$ are the Kronecker and Dirac delta functions respectively. $t$ and $t'$ denote different time instances, $k_B$ is the Boltzmann constant, and $T$ is the temperature. By integrating Eq.~(\ref{eq:Brown_space_time}) over the discretized time step $\Delta t$, we can derive the standard deviation of $\Vec{H}_{\text{th},i}$ in the $x, y, z$ directions:
\begin{align}
    \sigma_i = \frac{1}{\mu_0} \sqrt{\frac{2\alpha_i k_B T}{m_i \gamma (1+\alpha_i^2) \Delta t}}.
    \label{eq:H_th_sigma}
\end{align}
For each component $x, y, z$ of $\Vec{H}_{\text{th},i}$, we sample a value from a Gaussian distribution with a mean $\mu = 0$ and standard deviation given by $\sigma_i$ from Eq.~(\ref{eq:H_th_sigma}).

For the numerical integration, we use the Stratonovich interpretation of the stochastic process and implement Heun's method as our integration scheme. In the study by Ament {\em et al.}~\cite{ament_solving_2016}, Heun's integration scheme converges to the Stratonovich solution and preserves the norm of $\hat{m_i}$ for small time steps when simulating s-LLG dynamics.

\subsection{\label{sec:toffoli_implement}Implementation of Toffoli Gate Hamiltonian}

In our simulation, we assume identical parameters for all $n$ macrospins, allowing us to omit the index $i$ for the following quantities: $H_{Ai} = H_A$, $\alpha_i = \alpha$, and $m_i = m$. This approach models an array of perfectly identical macrospins. The simulation can then be simplified to focus solely on the evolution of the unit magnetic moment $\hat{m}_i$, as shown by the blue $\hat{m}$ trajectories in Fig.~\ref{fig:ising and macro}. The energy terms can thus be simplified as follows:
\begin{align}
    E_\text{ex} &= -\mu_0 m H_\text{ex}
    \sum_{\langle i,j \rangle}^{n} J_{ij} (\hat{m}_i \cdot \hat{m}_j),\label{eq:E_ex_simp} \\
    E_\text{anis} &= -\frac{1}{2}\mu_0 m H_{A}\sum_i^n  ( \hat{m}_i \cdot \hat{z} )^2, \label{eq:E_anis_simp} \\
    E_\text{Z} &=-\mu_0 m H_\text{ext} \sum_i^n h_i (\hat{m}_i\cdot \hat{z}). 
    \label{eq:E_Z_simp}
\end{align}

The Toffoli gate Hamiltonian $H_T$ can be directly mapped to the system's micromagnetic energy terms $E_\text{ex}$ and $E_\text{Z}$. Specifically, the $J_{ij}$ terms in $H_T$ in Eq.~(\ref{eq:H_Toffoli}) correspond to the $J_{ij}$ terms in $E_\text{ex}$ in Eq.~(\ref{eq:E_ex_simp}), and the $h_i$ terms in $H_T$ correspond to the $h_i$ terms in $E_Z$ in Eq.~(\ref{eq:E_Z_simp}). To maintain the relative strength between the coupling terms and the local field terms as given by $H_T$, we set the general scale of the exchange coupling strength $H_\text{ex}$ equal to the local field strength $H_\text{ext}$, \emph{i.e.}, $H_\text{ex} = H_\text{ext}$ for all trials. 

The ratio of anisotropy $H_A$ to coupling strength $H_\text{ex}$ influences the quality of the final output. High anisotropy causes spins to align more strictly along the north and south poles, emulating binary Ising spins. However, high anisotropy also makes it more difficult for spins to overcome the energy barrier to flip, increasing the likelihood that the system will be trapped in configurations that are not consistent with the truth table of the Toffoli gate. Conversely, if $H_A$ is relatively small, the spins can flip more easily, but this can result in the spins not aligning well with the $z$-axis, complicating the read-out process. In addition, thermal effects play a role in reaching the correct ground state when simulated annealing is incorporated. High temperatures produce thermal fluctuations that allow spins to overcome energy barriers and eventually find the ground state through annealing. 

Considering the many possible contributing factors, we simulate the zero-temperature dynamics and the dynamics with thermal annealing separately. First, we investigate the ability of pure (\emph{i.e.}, zero temperature) LLG dynamics to find the correct configurations given by the truth table as a function of $H_A/H_\text{ex}$, with $H_\text{ex}=H_\text{ext}$ fixed. The values of the parameters for our simulation are shown in Table~\ref{tab:params}. The values are realistic for the free layers of perpendicularly magnetized magnetic tunnel junction nanopillars~\cite{Kent2015}.

In addition, we study the ability of combined stochastic LLG dynamics with simulated annealing to reach the true ground state as a function of $H_A/H_\text{ex}$. We have empirically chosen an annealing schedule that decreases from $T = 300$ to $0\text{K}$ over $150$ evenly spaced steps. We set the exchange field $H_\text{ex}$ such that the exchange energy $E_\text{ex}$ is the order $~k_B T_\text{amb}$, with $T_\text{amb}=300$ K. This ensures the system has sufficient thermal energy to overcome energy barriers as it evolves. At each temperature, we let the system evolve with $10^5$ iterations, which corresponds to $1 \mu s$ in physical units. This time is longer than a typical spin relaxation time.  For instance, considering spin relaxation in the exchange field, $t_\text{relax}\approx 2 \pi/(\alpha \mu_0 \gamma H_\text{ex}) \approx 0.1 \mu s$, which is one order magnitude smaller than $1\mu s$. Therefore, our system will be in or very close to thermal equilibrium with the environment at the end of each temperature step.

\begin{table}[h]
\centering
\begin{tabular}{|c|c|c|}
\hline
Damping Constant & $\alpha$ & 0.01 \\
\hline
Magnetic Moment & m & $8\times10^{-19}$J/T \\
\hline
Ambient Temperature & $T_\text{amb}$ & 300K \\
\hline
Exchange Field & $H_\text{ex}$ &  $4.12\times10^3$ A/m \\
\hline
Simulation Time Step & $\Delta t$ & $0.01$ ns \\
\hline
Iterations at Each Temperature &   & $1 \times 10^5$ \\
\hline
Iterations at 0K (LLG) &   & $5 \times 10^5$ \\
\hline
\end{tabular}
\caption{Values of the parameters chosen for simulation. }
\label{tab:params}
\end{table}

\section{\label{sec:results}Results}

\subsection{\label{sec:0K}Zero temperature simulations}

We first simulate the construction of a Toffoli Gate using LLG dynamics at zero temperature ($T=0$ K). For a given initial configuration $(a,b,c,d,a',b',c')$, we begin by initializing the macrospins to the corresponding values $(\hat{m}_a,\hat{m}_b,\hat{m}_c,\hat{m}_d,\hat{m}_{a'},\hat{m}_{b'},\hat{m}_{c'})$. The input bits $(a,b,c)$ are converted into macrospins $(\hat{m}_{a},\hat{m}_{b},\hat{m}_{c})$ by aligning them strictly along the $\hat{z}$ axis, with bit 0 corresponding to $-\hat{z}$ and bit 1 to $+\hat{z}$. Then these three macrospins are numerically fixed throughout the simulation. We initialize the output bits $(d,a',b',c')$ as $(\hat{m}_d,\hat{m}_{a'},\hat{m}_{b'},\hat{m}_{c'})$ by first aligning these vectors along the $z$ axis. Then we add a small angular deviation of $0.01$ rad from $\hat{z}$ to prevent the spins from experiencing zero initial torque (\emph{i.e.}, to avoid the cross-product terms in Eq.~(\ref{eq:LLG}) being zero). 

\begin{figure}[h!]
 \centering\includegraphics[width=0.45\textwidth]{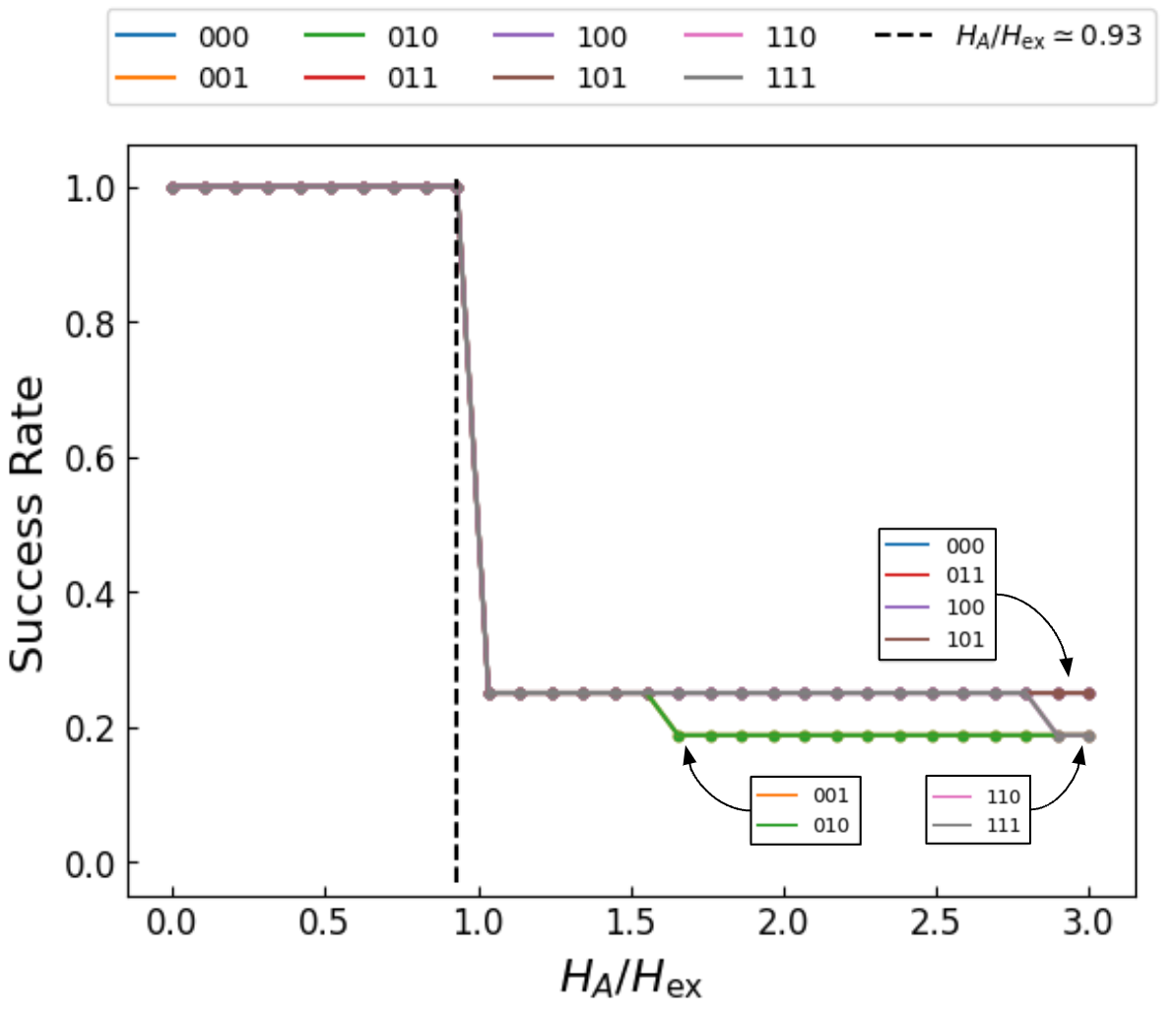}
 \caption{Success rate of the Toffoli gate reaching the correct output as a function of $H_A/H_\text{ex}$ ratio, under deterministic (zero temperature) LLG dynamics. The results for each input $(a,b,c)$ are color-coded in the plot. The graph appears to have fewer colors than listed due to multiple overlapping lines stacked on top of each other. The black dashed line at $H_A/H_\text{ex} \simeq 0.93$ represents the largest $H_A/H_\text{ex}$ in our simulation that still achieves a 100\% success rate for all trials.} 
 \label{fig:success rate 0K}
\end{figure}

\begin{figure}[h!]
 \centering
 \includegraphics[width=0.45\textwidth]{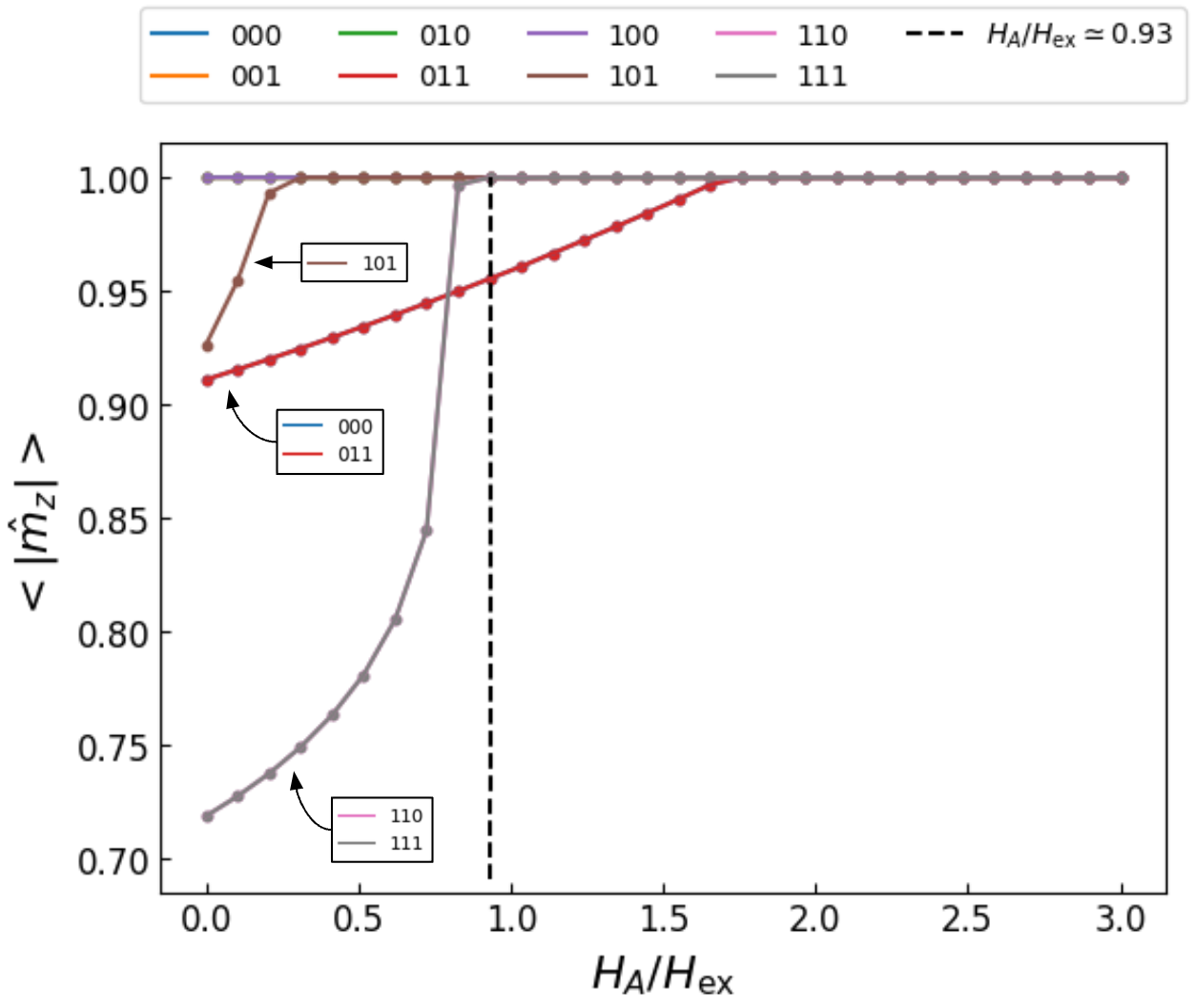}
 \caption{Average $\hat{z}$ projection of output spins $\langle | \hat{m}_z | \rangle$ as a function of $H_A/H_\text{ex}$ ratio, under deterministic (zero temperature) LLG dynamics. The results for each input configuration $(a,b,c)$ are color-coded in the plot. The graph appears to have fewer colors than listed due to multiple overlapping lines stacked on top of each other. The black dashed line at $H_A/H_\text{ex} \simeq 0.93$ represents the largest $H_A/H_\text{ex}$ in our simulation that still achieves a 100\% success rate for all trials.} 
 \label{fig:ising 0K}
\end{figure}

For each Toffoli Gate input $(a,b,c)$, there are 16 different possible initial configurations of $(d,a',b',c')$. For a fixed input $(a,b,c)$, we run 16 trials with different initial conditions of $(d,a',b',c')$, allowing the system to evolve to its final state. We record the final state and count the number of trials that result in an output $(a',b',c')$ consistent with the Toffoli gate truth table and use this to compute the success rate. The process is repeated for different $H_A/H_\text{ex}$ ratios, and the result is shown in Fig.~\ref{fig:success rate 0K}. We observe a $100\%$ success rate of reaching the correct configuration for $H_A/H_\text{ex}\lesssim 0.93$.

To assess how Ising-like the spins are at the end of the simulation, we record the absolute magnitude of the final $\hat{z}$ projection $ |\hat{m}_z|=|\hat{m} \cdot\hat{z}| $ for each output configuration $(\hat{m}_d,\hat{m}_{a'},\hat{m}_{b'},\hat{m}_{c'})$. There are 16 different trials for each input $(a,b,c)$, we take the average of $ | \hat{m}_z |$ across both the four output spins and the 16 trials to obtain a single value. The results are shown in Fig.~\ref{fig:ising 0K}. Since we set the external field applied locally to each macrospin as $\hat{e}_{\text{ext}} = \hat{z}$, an Ising-like output is still observed even when $H_A = 0$, with the smallest $\langle | \hat{m}_z | \rangle  = 0.72$. We also observe that as $H_A/H_\text{ex}$ increases, $\langle | \hat{m}_z | \rangle$ approaches 1, which indicates that the spins are more Ising like as the anisotropy increases.

\subsection{\label{sec:simulated_anneal}Simulation with thermal annealing (s-LLG)} 

\begin{figure}[h!]
 \centering\includegraphics[width=0.45\textwidth]{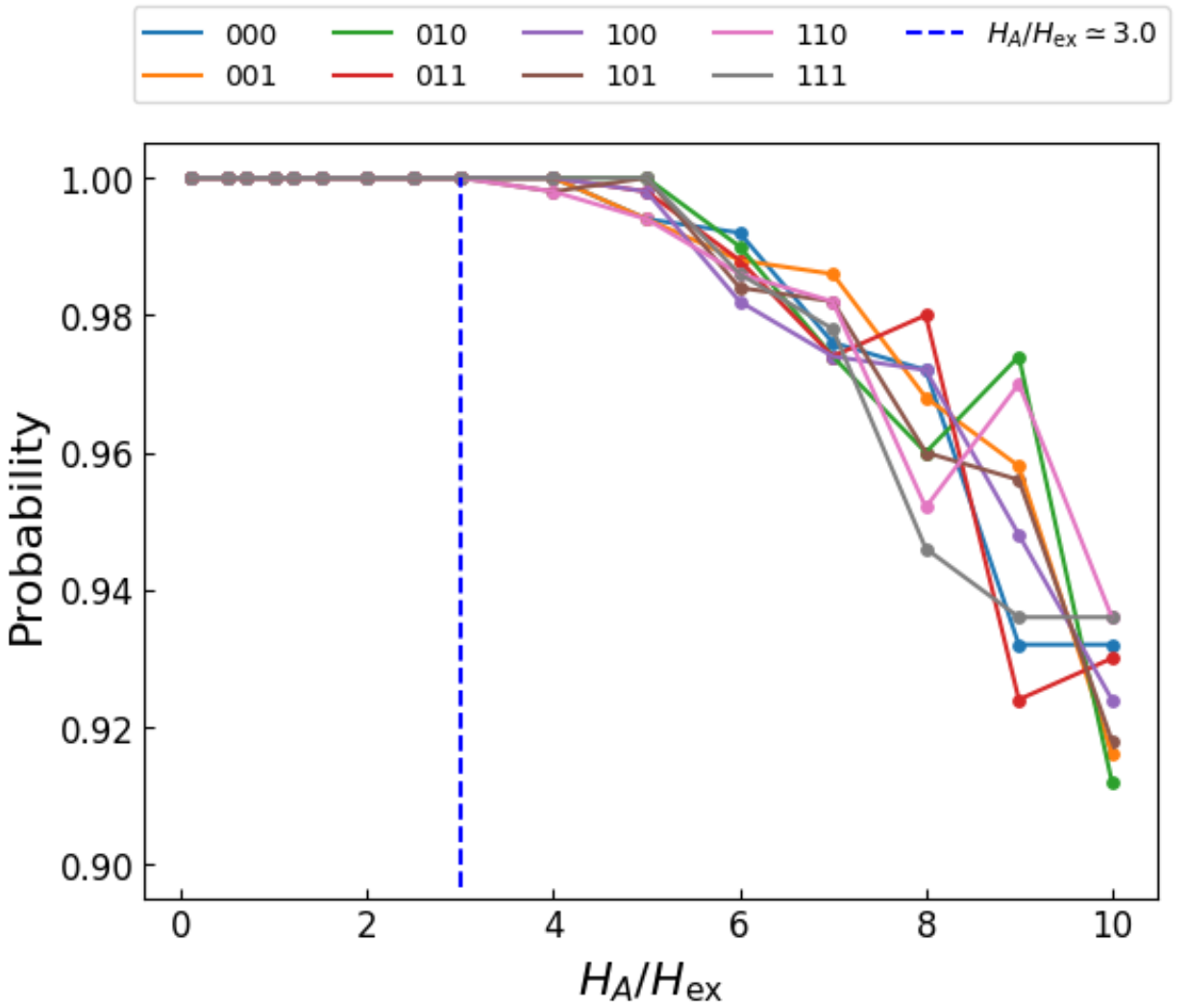}
 \caption{Probability of reaching the correct configurations at different $H_A/H_\text{ex}$ ratios for Toffoli gate construction with s-LLG dynamics and simulated annealing. The results for each $(a,b,c)$ are color-coded in the plot. For each $(a,b,c)$, the simulation is repeated for 500 times. Within the statistics of 500 trials, values above $H_A/H_\text{ex} \simeq 3.0$ fails to reach a 100\% probability of producing the correct output, as shown by the blue dashed line.} 
 \label{fig:success rate with anneal}
\end{figure}

\begin{figure}[h!]
 \centering
 \includegraphics[width=0.45\textwidth]{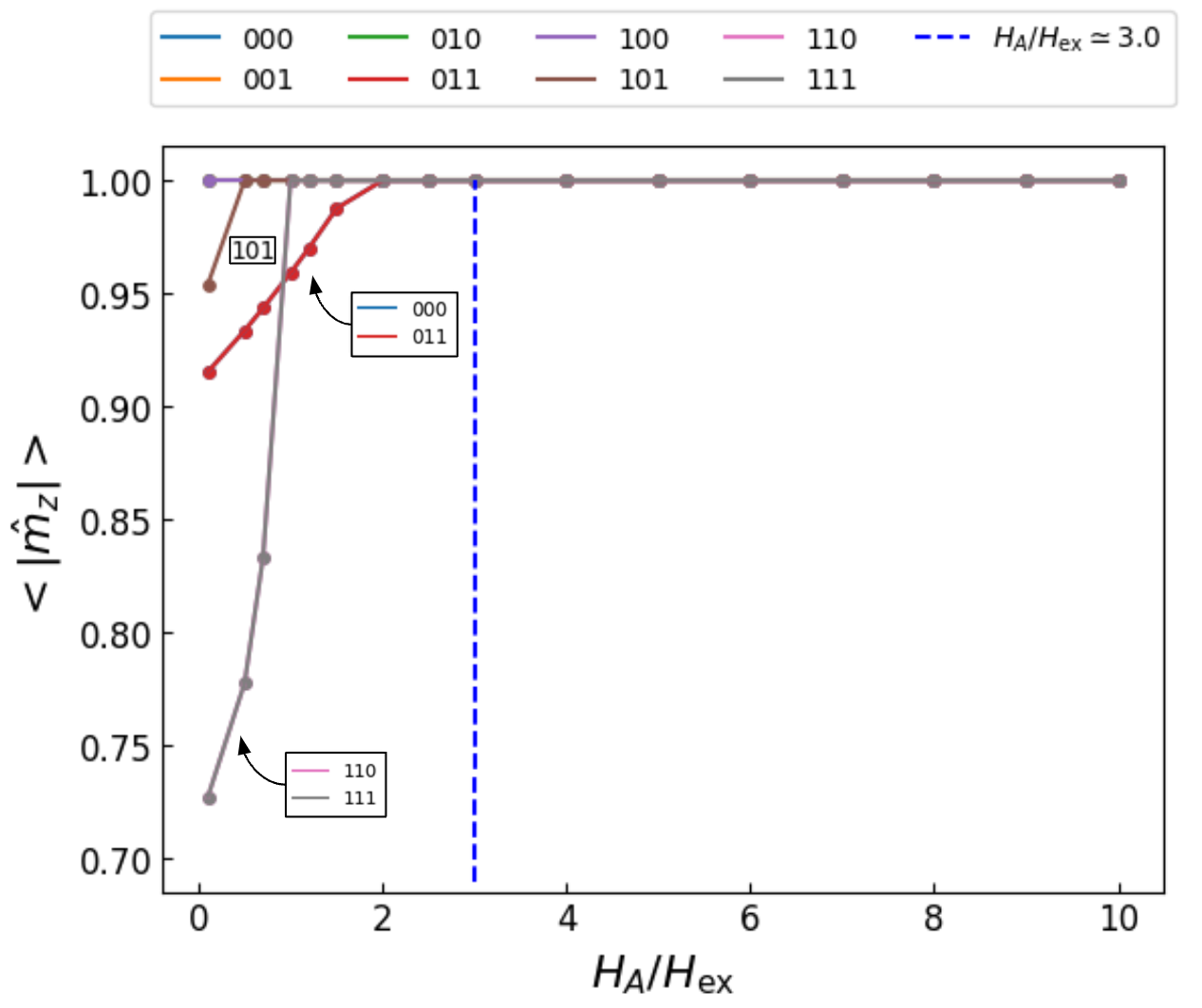}
 \caption{Average $\hat{z}$ projection of output spins, $\langle | \hat{m}_z | \rangle$, as a function of $H_A/H_\text{ex}$ ratios for Toffoli gate construction with s-LLG dynamics and simulated annealing. The average is taken across 500 trials. The results for each $(a,b,c)$ are color-coded in the plot. The blue dashed line at $H_A/H_\text{ex} \simeq 3.0$ represents the largest $H_A/H_\text{ex}$ allowed to achieve a 100\% probability of reaching the correct output with simulated annealing. } 
 \label{fig:ising with anneal}
\end{figure}

In this section, we simulate the Toffoli gate construction with simulated annealing. We fix the exchange coupling strength such that $E_\text{ex}$ between spins is on the order of $k_B T_\text{amb}$, where $T_\text{amb}$ is the ambient temperature of 300K. For each Toffoli Gate input $(a,b,c)$, we start with the configuration $(a,b,c,0,0,0,0)$ and perform an annealing process from 300 to 0K over 150 evenly spaced steps. Since we start at high temperature, the system is able to visit different magnetic configurations. Therefore, it is not as necessary to initialize the system with different configurations $(d,a',b',c')$, as in the 0K case. After the annealing process, we record the final state and repeat the simulations for 500 independent trials. We do the same simulations for different ratios $H_A/H_\text{ex}$. The results are shown in Figs.~\ref{fig:success rate with anneal} and~\ref{fig:ising with anneal}.

With the assistance of simulated annealing, we observe that within the statistics of 500 trials, the system is able to achieve a 100\% probability of finding the correct output for $H_A/H_\text{ex}$ up to  $H_A/H_\text{ex} \simeq 3.0$, which is higher than the limit of $H_A/H_\text{ex} \simeq 0.93$ observed with zero-temperature pure LLG dynamics. Further, at $H_A/H_\text{ex} \simeq 3.0$, the final configurations are all Ising-like, with all spins aligned along the $z$ axis as shown in Fig.~\ref{fig:ising with anneal}. 

\subsection{\label{sec:conclusion}Conclusion}

We have demonstrated a proof-of-concept Toffoli gate construction using macrospins with a uniaxial magnetic anisotropy that could be formed by the free magnetic layers of MTJs. Specifically, we investigated the influence of the ratio of the uniaxial magnetic anisotropy to the exchange coupling, $H_A/H_\text{ex}$, on the spin dynamics. Remarkably, the zero temperature LLG dynamics always ends in configurations consistent with the Toffoli gate's truth table provided that $H_A/H_\text{ex} \lesssim 0.93$. Introducing thermal annealing allows to push this ratio to higher values, up to $H_A/H_\text{ex} \simeq 3.0$, while keeping a 100\% success rate.

Our modeling thus shows that a universal reversible gate can be constructed from interacting classical macrospins, highlighting a promising avenue for research with arrays of interacting MTJs~\cite{talatchian_mutual_2021,phan_electrical_2022,schnitzspan_electrical_2023}. Although our simulation has shown the feasibility of this approach, further research is needed.  Specifically, this work explored the feasibility of constructing a single Toffoli gate. Future theoretical work could investigate the feasibility of building circuits using multiple Toffoli gates for real-world computational tasks. These would address the questions of methods, properties, and fidelity of connected Toffoli gates and, for example, how the time to solution scales with the problem size. 

\subsection*{Acknowledgement}
The research at NYU was supported by the Office of Naval Research (ONR) under Award No. N00014-23-1-2771 and in part by the National Science Foundation (NSF) under Award DMR-2105114. Research at the University of Lorraine was supported by ``MAT-PULSE,'' part of the French PIA project ``Lorraine Universit\'e d'Excellence'' reference ANR-15-IDEX-04-LUE. This work was also supported in part through the NYU IT High Performance Computing resources, services, and staff expertise. 

\newpage

\bibliography{mybib}
\end{document}